\documentclass[a4paper]{article}

\usepackage[english]{babel}
\usepackage[utf8x]{inputenc}
\usepackage[T1]{fontenc}

\usepackage[a4paper,top=3cm,bottom=2cm,left=3cm,right=3cm,marginparwidth=1.75cm]{geometry}

\usepackage{amsmath}
\usepackage{graphicx}
\usepackage[colorinlistoftodos]{todonotes}
\usepackage[colorlinks=true, allcolors=blue]{hyperref}

\title{Eliminating the effect of rating bias on reputation systems}
\author{Leilei Wu\textsuperscript{1},
Zhuoming Ren\textsuperscript{2},
Xiao-Long Ren\textsuperscript{3},
Jianlin Zhang\textsuperscript{2},
Linyuan L\"{u}\textsuperscript{1,2,*},
\\
$^{1}$ Institute of Fundamental and Frontier Sciences, University of Electronic Science 
\\ and Technology of China, Chengdu, P.R. China \\
$^{2}$  Alibaba Research Center for Complexity Sciences, Hangzhou Normal University, \\ Hangzhou, Zhejiang, P.R. China\\
$^{3}$  Computational Social Science, ETH Zurich, Zurich, Switzerland \\
$^{\dag}$  All the authors contributed equally to this work.\\
$^{*}$ linyuan.lv@gmail.com
}

\begin{document}
\maketitle

\begin{abstract}
The ongoing rapid development of the e-commercial and interest-base websites make it more pressing to evaluate objects' accurate quality before recommendation by employing an effective reputation system. The objects' quality are often calculated based on their historical information, such as selected records or rating scores, to help visitors to make decisions before watching, reading or buying. Usually high quality products obtain a higher average ratings than low quality products regardless of rating biases or errors. However many empirical cases demonstrate that consumers may be misled by rating scores added by unreliable users or deliberate tampering. In this case, users' reputation, i.e., the ability to rating trustily and precisely, make a big difference during the evaluating process. Thus, one of the main challenges in designing reputation systems is eliminating the effects of users' rating bias on the evaluation results. To give an objective evaluation of each user's reputation and uncover an object's intrinsic quality, we propose an iterative balance (IB) method to correct users' rating biases. Experiments on two online video-provided Web sites, namely MovieLens and Netflix datasets, show that the IB method is a highly self-consistent and robust algorithm and it can accurately quantify movies' actual quality and users' stability of rating. Compared with existing methods, the IB method has higher ability to find the ``dark horses", i.e., not so popular yet good movies, in the Academy Awards.
\end{abstract}

\section{Introduction}
The fast development of the Internet and related infrastructures has created vast opportunities for people to date, read, shop, and enjoy entertainment online~\cite{watts2007twenty, vespignani2009predicting, RuiMao2015}. As people come to rely more and more on the Internet, they place themselves at additional risk. Disinformation and rumors mislead people into making wrong decisions. For example, some e-commercial Web sites sellers manipulate information in order to present low quality products in a good light. How to effectively disentangle truth from falsehood to protect individuals from malicious deception is a critical problem, especially for the companies who provide information services or products online~\cite{Alter2001, lu2012recommender, ren2014avoiding, liaohao2015}. Reputation systems arose as a result of the need for Internet users to gain trust in the individuals they transact with online~\cite{Resnick:2000:RS:355112.355122, liny2016PR}. Additionally, reputation systems enable users and customers to better understand the provided information, products, and services~\cite{sun2012security,liao2014Network}. Reputation systems may help users to make decisions on whether or not to buy specific services or goods that they have no prior experience using or never purchased before~\cite{shang2010empirical,ni2014ceiling,liu2013empirical}.

Reputation system use a collection of historical ratings records and attributes of users' and items' to calculate their reputation/quality levels, which usually represented as the form of scores. 
Most e-commercial and interest-based websites employed some kinds of reputation systems to differentiate the qualities of services, products or entities before recommendation or information push. For example, Netflix, which provides DVD rental service allows users to vote on the movies and then computes the reputation score of each movies. Since the ratings have a large influence on users' online purchasing decisions and the online digital content distribution, various algorithms have been proposed to give objective evaluations. Laureti \textit{et al.}~\cite{laureti2006information} proposed an iterative refinement (IR) method where a user's reputation, i.e., rating stability is inversely proportional to the difference between the user's ratings and the corresponding objects' estimated quality. The estimated quality of each object and reputation of each user are iteratively updated until convergence is reached. Zhou \textit{et al.}~\cite{zhou2011robust} proposed a robust ranking algorithm where a user's reputation is calculated by the Pearson correlation between user's ratings and the objects' estimated quality. Compared with the IR method, this method shows a higher robustness against spammer attacks. More recently, Liao \textit{et al.}~\cite{liao2014ranking} developed a reputation redistribution process to the iterative ranking measurement, which can effectively increase the weight of votes cast by highly reputable users and reduce the weight of users with a low reputation, when estimating the quality of objects. There are also some other algorithms that are built on the basis of bayesian theory~\cite{whitby2004filtering}, belief theory~\cite{Yu:2002:EMD:544741.544809}, the flow model~\cite{brin1998anatomy}, or fuzzy logic concepts~\cite{bharadwaj2009fuzzy}. Most of the previous methods are based directly on ratings while neglect the fact that users may have a personal bias when they give a score to an object. We have empirically investigated four benchmark datasets that were obtained from two video-provided Web sites, MovieLens~\cite{MovieLens} and Netflix~\cite{Netflix} and found that each user has a certain magnitude of rating error which decreases the prediction accuracy of ratings~\cite{ochi2012rating}. In order to eliminate the effects of this rating error on the evaluation results, we propose a new algorithm called the iterative balance (IB) method. Experiments on MovieLens and Netflix datasets show that the IB method is a highly self-consistent and robust algorithm, it can accurately quantify a user's reputation and a movie's quality. Compared with the state-of-the-arts, the IB method has a greater ability to find the ``dark horses'' for Oscar award.

This paper is organized as follows. In section 2, we introduce the representation of rating systems and the general framework of iterative ranking algorithms. Next, we describe our IB method and some well-known iterative algorithms which will be used for comparisons. In section 3, four benchmark datasets and several evaluation metrics are described. In section 4, we show the performance of the IB method in terms of accuracy and robustness. Conclusions and discussions are drawn in the last section where the potential relevance and applications of the IB method are discussed.

\section{Materials and Methods}
\subsection{Bipartite network representation of rating systems}
\begin{figure}[ht]
\center\scalebox{0.8}[0.8]{\rotatebox{0}{\includegraphics{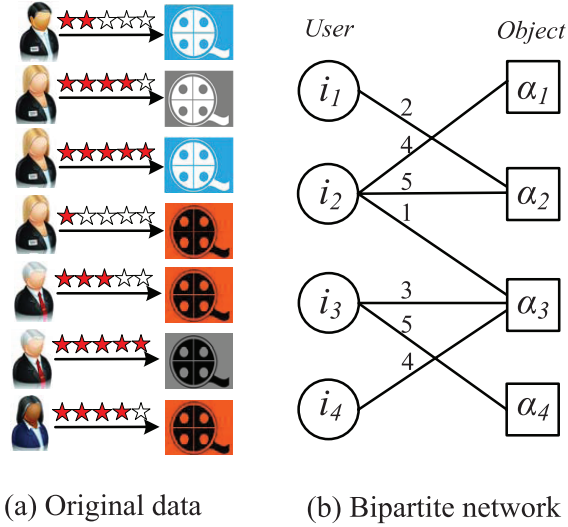}}}
\caption{\textbf{An example of how to construct a user-object bipartite network based on a collection of rating data.} (a) is a real scene in which users see movies and vote them in five discrete ratings 1-5. (b) is the corresponding bipartite network, in which users and objects are represented by circles and squares, respectively.}\label{Fig1}
\end{figure}

Bipartite networks are commonly used to represent the relationships between two groups of entities, such as the relationships between actors and movies, goods and customers, books and readers, publications and authors, etc. Only the relationships between the two groups of entities are allowed. Here, we use bipartite networks to represent the rating systems which include the set of users (denoted by $U$), the set of objects (denoted by $O$) and the ratings between users and objects (denoted by $R$). A link in the bipartite network connecting user~$i$ and object~$\alpha$ represents a historical rating $r_{i\alpha}$ ($\in R$). We give a simple example in Fig.~\ref{Fig1} to show how to construct a bipartite network based on a set of rating data. The original data shown in Fig.~\ref{Fig1}(a) has seven rating records made by four users on four movies. The ratings are given on the integer scale from 1 star to 5 stars (i.e., worst to best). Fig.~\ref{Fig1}(b) shows the corresponding bipartite network where users are represented by circles, and objects are presented by squares. Users are connected with the movies that they have rated. All the users who have rated object~$\alpha$ are represented by set $U_\alpha$, while all the objects which have been rated by user $i$ are represented by set $O_i$. For example, in Fig.~\ref{Fig1} $U_{\alpha_3}=\{i_2,i_3,i_4\}$ and $O_{i_2}=\{\alpha_1,\alpha_2,\alpha_3\}$. The object $\alpha$'s degree~$k_{\alpha}$ is the number of users in set~$U_\alpha$, and the user $i$'s degree~$k_{i}$ is the number of objects in the set~$O_i$.

\subsection{Iterative ranking framework}
As a matter of fact, items have a set of qualities, based on a set of $N$ traits. A user’s aggregate rating is a reflection of the quality of those traits, plus the individual weighting that reflects the user’s value system. A user’s reputation is the accuracy of rating those traits, independent of his individual weighting of the traits. For convenience, This paper deals with the case where $N=1$. 
$Q_{\alpha}$ and $R_{i}$ denote the quality of object $\alpha$ and the reputation of user $i$, respectively. Note that, when users' biases and mistakes are absent, i.e., $R_{i}=1$ for every user, any two users would rate any object the same according to the instinct quality of the object.
The most straightforward method to quantify one object's quality is to consider the historical ratings that the object received. Averaging over all ratings (abbreviated as AR) is the simplest method, which mathematically reads
\begin{equation}
\bar{Q}_{\alpha}=\frac{\sum\limits_{i\in U_{\alpha}}r_{i\alpha}}{k_\alpha}.
\end{equation}
Obviously, in this form the ratings from different user contribute equally to $\bar{Q}_{\alpha}$. However, the ratings of users with higher reputation are more reliable than the ratings from low reputation users. Therefore a weighted form to calculate the quality of an object $Q_{\alpha}$ was proposed.
\begin{equation}
Q_{\alpha}=\sum\limits_{i\in U_{\alpha}}R_{i}r_{i\alpha},
\label{QAlpha}
\end{equation}
where $R_i$ is usually the normalized reputation score of user $i$.

KR is ``a very crude approach" of evaluating the reputation of users in system. The basic assumption is that the user with more experience, i.e., rating more items before, has a higher ability to rating trustily and precisely. The reputation of a user is directly proportional to the number of items he or she has rated in KR. However, due to the unreliable of this assumption, nothing more will be discussed in this paper.

There are also three iterative ways to calculate each user's reputation score $R_i$. Laureti \textit{et al.}~\cite{laureti2006information} presented an iterative refinement method (abbreviated as IR), which considers users' reputation scores as inversely proportional to the mean squared error between users' rating records and the quality of objects, namely
\begin{equation}
TR_{i}=\frac{k_i}{\sum\limits_{\alpha\in{O_i}}(Q_\alpha-r_{i\alpha})^{2}},\label{1}
\end{equation}
After normalization, we obtain
\begin{equation}
IR_{i}=\frac{TR_i}{\sum\limits_{j\in U}TR_j}.\label{2}
\end{equation}

Zhou \textit{et al.} \cite{zhou2011robust} proposed a correlation-based iterative method (abbreviated as CR), which assumes that a user's reputation is calculated by the Pearson correlation \cite{pearson1895note} between user rating records and the corresponding objects' quality.
\begin{equation}
corr_i=\frac{k_i\sum\limits_{\alpha\in{O_i}}r_{i\alpha}Q_{\alpha}-
\sum\limits_{\alpha\in{O_i}}r_{i\alpha}\sum\limits_{\alpha\in{O_i}}Q_{\alpha}}{\sqrt{k_i\sum\limits_{\alpha\in{O_i}} r^2_{i\alpha}-(\sum\limits_{\alpha\in{O_i}}r_{i\alpha})^2}\sqrt{ k_i\sum\limits_{\alpha\in{O_i}}Q^2_{\alpha}-(\sum\limits_{\alpha\in{O_i}}Q_{\alpha})^2}},\label{5}
\end{equation}
The reputation scores are defined as
\begin{equation}
TR_{i}=\left\{
\begin{array}{cc}
corr_i&\textrm{if $corr_i\geq0$,}\\
0&\textrm{if $corr_i<0$.}
\end{array}
\right.
\end{equation}
Normalizing $TR_i$, we obtain
\begin{equation}
CR_{i}=\frac{TR_i}{\sum\limits_{j\in U}TR_j}.\label{CR}
\end{equation}

More recently, Liao \textit{et al.} \cite{liao2014ranking} proposed a reputation redistribution process (abbreviated as IARR) to improve the validity by enhancing the influence of highly reputed users. Then equation~(\ref{CR}) can be rewritten as
\begin{equation}
IARR_{i}=TR^\theta_i\frac{\sum\limits_{j\in U}TR_j}{\sum\limits_{j\in U}TR^\theta_j}.
\label{IARR}
\end{equation}
where $\theta$ is a tunable parameter to control the influence of reputation. Obviously, 
when $\theta=0$, $IARR_i$ is a constant value for all the users;
when $\theta=1$, IARR reduces to the CR method. In this paper, we set $\theta=3$ which is suggested by the proposers ~\cite{liao2014ranking}. In the same time, Liao \textit{et al.} also presented another similar algorithm, called IARR2, by introducing a penalty factor to IARR. IARR2 algorithm thought that a user is more reliable if he rates more objects and his reputation is still high, and so does the objects. In IARR2, the equation~(\ref{QAlpha}) should be written as

\begin{equation}
Q_{\alpha}=\max_{i\in U_{\alpha}}\{IARR_{i}\}\sum\limits_{i\in U_{\alpha}}IARR_{i}r_{i\alpha}
\label{IARR2Q}
\end{equation}
and the $TR_{i}$ in equation~(\ref{IARR}) was revised as
\begin{equation}
TR_{i}=\left\{
\begin{array}{cc}
\frac{\lg(k_{i})}{\max\{\lg(k_{i})\}}\cdot{corr_i}&\textrm{if $corr_i\geq0$,}\\
0&\textrm{if $corr_i<0$.}
\end{array}
\right.
\label{IARR2R}
\end{equation}
In summary, under the framework of iterative models, there are four steps to achieve the final results through four different algorithms:

(\romannumeral1) Initialize the reputation of users. Specifically, we set $IR_i(0)=1/|O|$, $CR_i(0)=k_i/|O|$, $IARR_i(0)=k_i/|O|$ and $IARR2_i(0)=k_i/|O|$ for the IR, CR, IARR and IARR2 methods, respectively \footnote{We have checked the results when the initialization of IR is the same as the other three algorithms, i.e., $IR_i(0)=k_i/|O|$. The results are exactly the same as the case when $IR_i(0)=1/|O|$. To follow the original paper of IR method, we use $IR_i(0)=1/|O|$ in our paper and experiments.}. 

(\romannumeral2) Estimate the quality of each object with equation (\ref{QAlpha}), where $R_i$ can be $IR_i$ (equation~(\ref{2})), $CR_i$ (equation~(\ref{CR})) and $IARR_i$ (equation~(\ref{IARR})), while IARR2 can be calculated based on IARR according to equation~(\ref{IARR2Q}).

(\romannumeral3) Update the reputation of each user according to equations~(\ref{1})(\ref{2}) for IR, equations~(\ref{5})-(\ref{CR}) for CR, equation~(\ref{IARR}) for IARR methods, and equation~(\ref{IARR2R}) and equation~(\ref{IARR}) for IARR2, respectively.

(\romannumeral4) Continue the iteration process according to (\romannumeral2) and (\romannumeral3) until the change of the quality estimates $\sum\limits_{j\in O}(Q_\alpha(t)-Q_\alpha(t-1))$ is less than a threshold $\varepsilon$, then terminate the iteration. In our experiments, we set $\varepsilon=10^{-6}$.

\subsection{Iterative balance model}
The above three methods neglect the fact that the ratings of different users may have bias due to personal interests and criteria. This bias can be measured by the standard deviation and the skewness of the user's rating records. Let's consider $|U|$ users and $|O|$ objects. Each user $i$ has a certain magnitude of rating error $\delta_i$ and each object $\alpha$ has an intrinsic quality $Q_\alpha$ which is unknown for users. The magnitude of rating error 
$\delta$ indicates the inaccuracy degree of the rating score, which could play negative or positive effect on the rating.
Then the rating of user $i$ on object $\alpha$, namely $r_{i\alpha}$ can be written as
\begin{equation}
r_{i\alpha}=Q_{\alpha}+\delta_{i}.
\end{equation}
Here, we assume that the distribution of the magnitude of rating error $\delta$ has zero mean.
For an arbitrary user $i$, his magnitude $\delta_i$ can be measured by the standard deviation ($SD$), which reads
\begin{equation}
SD_{i}=|\delta_{i}-\langle \delta \rangle|=\sqrt{\frac{\sum\limits_{\alpha\in{O_i}}(r_{i\alpha}-\bar{r}_{\alpha})^2}{k_i}},
\end{equation}
where $\bar{r}_{\alpha}$ is the average score of all ratings on object $\alpha$. Furthermore, we also give the skewness of the rating records, which refers to asymmetry in the real distribution of a user's rating records about its mean,
\begin{equation}
SK_{i}=\frac{k_i\sum\limits_{\alpha\in{O_i}}(r_{i\alpha}-\bar{r}_{\alpha})^3}{(k_i-1)(k_i-2)SD^3_{i}},
\end{equation}
where $SK_{i}$ could come in the form of `negative skewness' or `positive skewness', depending on whether the user's rating records are skewed to the left (negative skew) or to the right (positive skew) of the average rating records.

\begin{figure}[ht]
\center\scalebox{0.55}[0.55]{\rotatebox{270}{\includegraphics{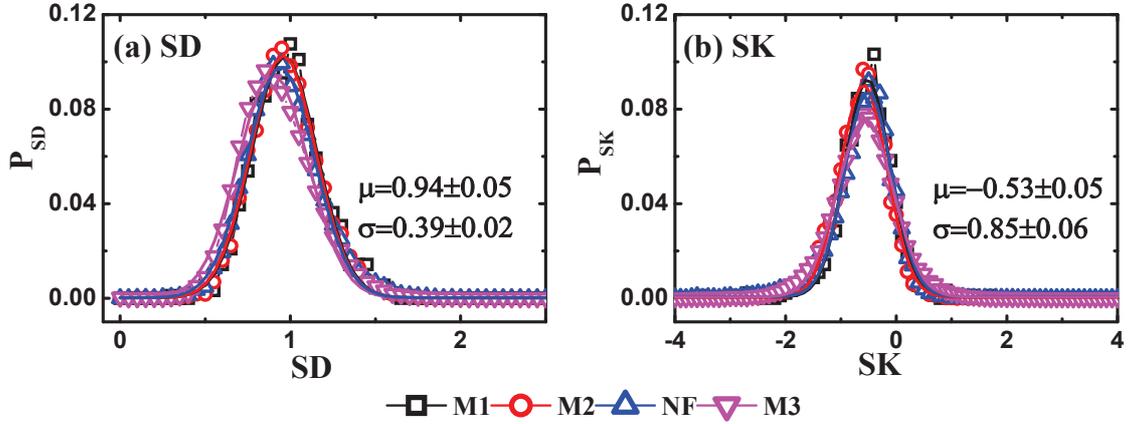}}}
\caption{\textbf{Distribution of the users' rating magnitude in the M1, M2, M3, and NF datasets.} (a) the distribution of users' standard deviation ($SD$) and (b) the distribution of users' skewness ($SK$) in the four datasets. The statistical features of the four datasets are shown in Table 1. Detailed introduction of the datasets can be found in section \textit{Materials and Methods}.}\label{Fig2}
\end{figure}

We empirically analyze four benchmark user-movie datasets, three of them are samples from MovieLens, named M1, M2 and M3, and the other one is from Netflix, named NF (see Table 1 for basis statistics of the datasets). For each dataset, we investigate the distribution of $SD$ and $SK$ of users, respectively shown in Fig.~\ref{Fig2}(a) and Fig.~\ref{Fig2}(b). Both $SD$ and $SK$ follow normal distribution where the parameters are estimated via maximum likelihood approximation method. Due to the user's personal bias of rating, we proposed an iterative balance model to eliminate the bias in order to better quantify the user's reputation. The model considers the user magnitude to meet equation (9), and its process can be described as follows:

(\romannumeral1) Initialize the quality of each object according to equation (1), we obtain $Q_{\alpha}(0)=\bar{Q}_{\alpha}$.

(\romannumeral2) Update the reputation of each user according to
\begin{equation}
%IBR_{i}(t)=\frac{\sqrt{\sum\limits_{\alpha\in{O_i}}[Q_{\alpha}(t-1)-r_{i\alpha}]^2}}{k_i}.
IBR_{i}(t)=\sqrt{\frac{\sum\limits_{\alpha\in{O_i}}[Q_{\alpha}(t-1)-r_{i\alpha}]^2}{k_i}}.
\end{equation}
$IBR_{i}$ measures the rating bias of user $i$. Obviously, the lower the $IBR_{i}$ is, the higher reputation the user $i$ has.

(\romannumeral3) Update the quality of each object according to the equation
\begin{equation}
Q_{\alpha}(t)=\frac{1}{k_\alpha}\sum\limits_{i\in U_{\alpha}}[r_{i\alpha}+IBR_{i}(t)sgn(Q_{\alpha}(t-1)-r_{i\alpha})],
\end{equation}
where $sgn(x)$ is the sign function, which returns $1$ if $x>0$; $-1$ if $x<0$; and $0$ for $x=0$. It is noted that if $Q_{\alpha}(t)<0$, then $Q_{\alpha}(t)=0$.

(\romannumeral4) Continue the iteration process of (\romannumeral2) and (\romannumeral3) until the change of the quality estimate $\sum\limits_{j\in O}(Q_\alpha(t)-Q_\alpha(t-1))$ is less than a threshold $\varepsilon=10^{-6}$, then terminate the iteration. The final stable values of $Q_{\alpha}(t_c)$ and $IBR_i(t_c)$ are used to quantify the intrinsic quality of object $\alpha$ and the reputation of user $i$, respectively.

\section{Data and Metric}
\subsection{Datasets}
To test the performance of our IB method, we consider four benchmark datasets, which are sampled from MovieLens~\cite{MovieLens} and Netflix~\cite{Netflix}. MovieLens is an online movie recommendation Web site, who invites users to rate movies. Netflix Web site also has DVD rental service and the users can vote on the movies. The first three datasets are sampled from MovieLens with different sizes, which are named as M1, M2 and M3. The fourth dataset is a random sample of the whole records of user activities on Netflix.com. The rating scale for both MovieLens and Netflix is from one (i.e., worst) to five (i.e., best). Based on the users' historical records, we can construct a user-movie bipartite network. If user $i$ selects movie $\alpha$ and rates it, a link between user $i$ and movie $\alpha$ would be established. The statistical features of the four networks constructed based on four datasets are summarized in Table~\ref{table1}. In this paper, we consider only users and objects with degrees greater than $20$.

\begin{table}[!ht]
\centering
\caption{\textbf{The statistical features of the four real datasets.} $|U|$, $|O|$ and $|R|$ are the number of users, objects and rating records (i.e. the number of edges of bipartite networks), respectively. $<k_{\alpha}>$ and $<k_i>$ are the average degree of objects and users, respectively.}
\begin{tabular}{crrrccc}
\hline
\bf{Data Sets}  & \bf{$|U|$}       & \bf{$|O|$}       & \bf{$|R|$}    & \bf{$<k_{\alpha}>$} & \bf{$<k_i>$}     & \bf{sparsity} \\ \hline
$M1$ & 943     & 1,682   & 100,000     & 60          & 106   & 0.0630    \\ 
$M2$ & 6,040   & 3,952   & 1,000,209   & 253         & 166   & 0.0419    \\ 
$M3$ & 10,681  & 69,878  & 10,000,054  & 936         & 143   & 0.0134    \\ 
$NF$ & 10,000  & 6,000   & 824,802     & 137         & 82    & 0.0137    \\ \hline
\end{tabular} \label{table1}
\end{table}

\subsection{Evaluation metrics}
To evaluate the performance of IB method, we employ the mean-squared error (MSE) to measure the algorithm's accuracy on quantifying users' reputation, and the precision to evaluate the algorithm's accuracy on identifying good movies. Besides accuracy, we also investigate the robustness of our method, which is measured by the MSE and the Kendall's tau~($\tau$) coefficient~\cite{nelson2001kendall}.

A good method should give a higher reputation score to users with a lower error magnitude. $MSE(i)$ represents the scoring stability of user $i$, which reads
\begin{equation}
MSE(i)=\frac{\sum\limits_{\alpha\in O_i}(r_{i\alpha}-Q_\alpha)^2}{k_i}
\end{equation}
where $Q_\alpha$ is the intrinsic quality of object $\alpha$, i.e. the final quality value $Q_\alpha(t_c)$. Usually, the comparisons focus on the top-rank users, therefore we here consider the average MSE value of the $L$ highest reputation users.
\begin{equation}\label{MSE}
MSE=\frac{\sum MSE(i)}{L}.
\end{equation}
Lower MSE value indicates higher accuracy.

The accuracy of measuring object quality is evaluated by comparing with the movies nominated at Annual Academy Awards \cite{Oscar} and Golden Globe Awards \cite{Golden}. These nominated movies are the benchmark good movies in the evaluation. A good algorithm will rank the benchmark movies higher than others, therefore we apply \textit{precision} to evaluate the ability of an algorithm to find good movies. Instead of considering all movies, we focus on the top-$L$ places. Then precision is defined as
\begin{equation}
P(L)=\frac{m}{L},
\end{equation}
where~$m$ indicates the number of benchmark movies existing in the top-$L$ places of the ranking list. Higher precision corresponds with better performance.

The robustness is measured by Kendall's tau ($\tau$) coefficient~\cite{nelson2001kendall}. For a dataset, each method gives a ranked list of objects. If movie $A$ is better than movies $B$ in dataset M1, then a robust algorithm will also rank movie $A$ higher than movie $B$ in dataset M2 (or M3). To measure the robustness, we consider the common objects in two datasets (i.e., M1 and M2, M1 and M3, M2 and M3), and extract the sub-ranking list of the common objects from each original ranked list. Assume there are $N$ common objects between two lists where the quality score of object $i$ are denoted by $Q_i$ and $Q'_i$, respectively. The Kendall's tau rank correlation coefficient counts the difference between the number of concordant pairs and the number of discordant pairs, which reads
\begin{equation}
\tau=\frac{\sum\limits_{i=1}^{N}\sum\limits_{j=1}^{N}sgn[(Q_i-Q_j)(Q'_i-Q'_j)]}{N(N-1)},
\end{equation}
where $sgn(x)$ is the sign function, which returns $1$ if $x>0$; $-1$ if $x<0$; and $0$ for $x=0$. Here $(Q_i-Q_j)(Q'_i-Q'_j)>0$ means concordant, and negative value means discordant. The higher $\tau$ value is, the more robust the algorithm is. In the ideal case, $\tau=1$ indicates that the two ranking lists are exactly the same.

\section{Results}
\subsection{Accuracy for quantifying users' reputation}
\begin{figure}[ht]
\center\scalebox{0.5}[0.5]{\rotatebox{270}{\includegraphics{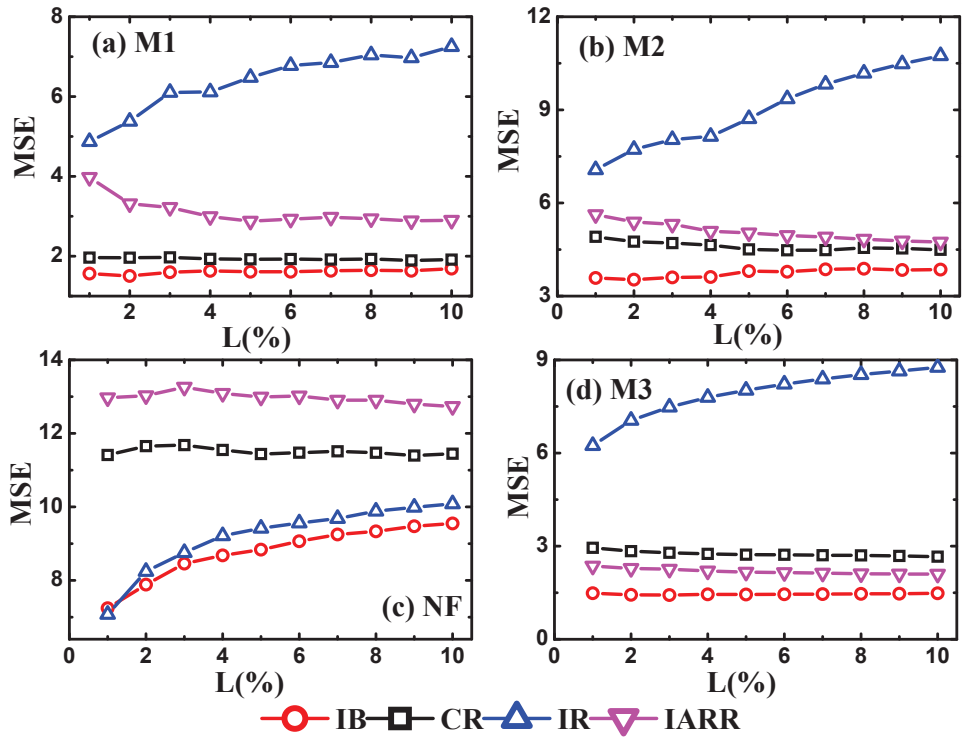}}}
\caption{\textbf{Accuracy of algorithms for quantifying users' reputation measured by MSE.} $L$ equals the percent [1\%,10\%] of $|U|$. The four methods include the IB, CR, IR and IARR.}\label{Fig3}
\end{figure}

Fig.~\ref{Fig3} shows the MSE value for IB method with different $L$, see equation~(\ref{MSE}). We also present other representative algorithms for comparisons. 
However, the penalty factor in IARR2 amplifies the value of the users' reputation and objects' quality greatly, which makes the MSE value of IARR2 much bigger than other methods. If we plot the curve of IARR2 in Fig.~\ref{Fig3}, all the other curves will become nearly linear. So the the MSE result for IARR2 is not present here. 
We could observe that as $L$ increases, the MSE value of the IB method is always the lowest, indicating that the IB method is a good measure of quantifying user reputation. Besides, we also investigate the correlation between the users' reputation scores and their personal MSE values. Table ~\ref{Utau} shows the Kendall's tau correlation coefficient between the two ranking lists respectively generated by ranking users decreasingly according to their reputation scores (the higher the better) and by ranking users increasingly according to their MSE values (the lower the better). For all four datasets, the IB method yields the highest value, indicating that our IB method is highly self-consistent.

\begin{table}[!ht]
\centering
\caption{\textbf{The correlation between the users' reputation scores and their personal MSE values.} We apply Kendall's tau to measure the correlation between the two ranking lists respectively generated by ranking users decreasingly according to their reputation scores (the higher the better) and by ranking users increasingly according to their MSE values (the lower the better). For each dataset, the highest value is emphasized in bold.}
\begin{tabular}{ccccc}
\hline
\bf{Data Sets} & \bf{IB}      & \bf{CR} & \bf{IR}   & \bf{IARR}   \\ \hline
$M1$         & \textbf{0.980} & 0.433   & 0.945     & 0.192  \\ 
$M2$         & \textbf{0.664} & 0.169   & 0.638     & 0.213  \\ 
$M3$         & \textbf{0.371} & 0.146   & 0.279     & 0.247  \\ 
$NF$         & \textbf{0.449} & 0.115   & 0.311     & 0.255  \\ \hline
\end{tabular} \label{Utau}
\end{table}

\subsection{Accuracy for identifying good objects}
Firstly, how do you define good objects? More specifically, how do you define good films? This is a well-known and highly controversial issue so that the opinion concerning this topic varies from person to person. According to a collection of answers in Quora.com, many people define a good films by how much it entertains and/or moves audience, how much it related to audience, or how strongly it makes audience emote. Just as the saying goes, "Each reader creates his own Hamlet". Here we want to adopt the movies that are most interesting, most appealing and most exciting as the benchmarks of the good films, and we believe that the selecting of movies that were nominated by either the Academy Awards~\cite{Oscar} or the Golden Globe Awards~\cite{Golden} should be an authority choice. We adopt the precision to calculate the accuracy for identifying good movies. In table~\ref{nominate}, we summarize the number of nominated movies in three MovieLens datasets. 

Note that, users' behaviors in the movies rating website changes over time, particularly before and after a movie be awarded in famous film festival like Academy Awards or Golden Globe Awards. 
The Academy Awards was first presented in 1929 while Golden Globe Awards was first presented in 1943. 
However, the two data sets, Netflix and Movielens, we used in our manuscript are created in recent decades. This means that most of the rating scores are created after the movies were awarded in the film festival. The data sets we obtained limited us to explore the rating dynamics over time in this paper. We will try to study this problem in our future works.

\begin{figure}[ht]
\center\scalebox{0.65}[0.65]{\rotatebox{270}{\includegraphics{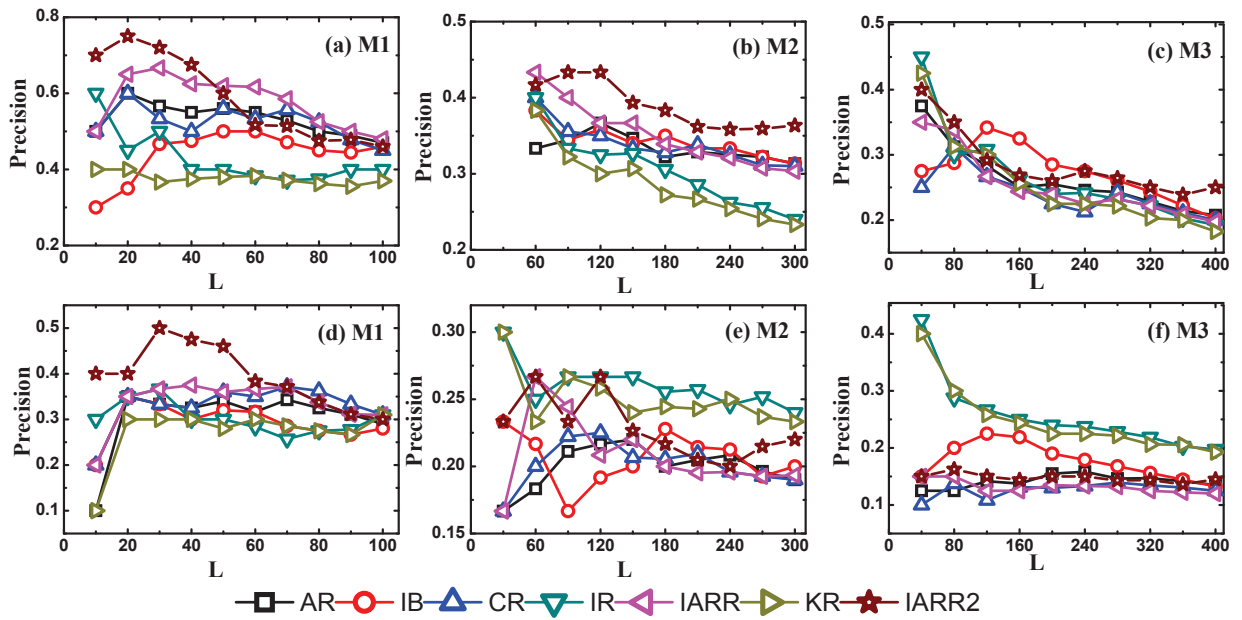}}}
\caption{\textbf{Algorithmic accuracy for identifying good movies measured by $precision$.} Due to the different size of the three datasets, we set the maximum $L$ equal to $100$, $300$ and $400$ for M1, M2 and M3, respectively. (a)-(c) are results found by evaluation with the Academy Awards benchmarks, while (d)-(f) are results found by evaluation with the Golden Globe Awards benchmarks.}\label{Fig4}
\end{figure}

Fig.~\ref{Fig4} shows the precision of five methods, including the IB, AR, CR, IR and IARR methods, on identifying good movies. For all methods, the precision decreases with the increase of $L$. Generally speaking, our IB method does averagely well. In some cases, IB performs good. For example, in M3 dataset the IB performs the best when evaluating with the Academy Awards, but is defeated by IR method when evaluating with the Golden Globe Awards.

\begin{table}[!ht]
\centering
\caption{\textbf{The number of nominated movies in the MovieLens datasets.} Both the Academy Awards and the Golden Globe Awards are considered. The numbers in brackets are the corresponding average number of ratings.}
\begin{tabular}{llll}
\hline
\bf{Award type}       & \bf{M1}       & \bf{M2}   & \bf{M3}    \\ \hline
$Academy Awards$      & 108 (175)     & 221 (717) & 361 (3942) \\ 
$Golden Globe Awards$ & 98            & 220       & 372        \\  \hline
\end{tabular}\label{nominate}
\end{table}

\begin{figure}[ht]
\center\scalebox{0.65}[0.65]{\rotatebox{270}{\includegraphics{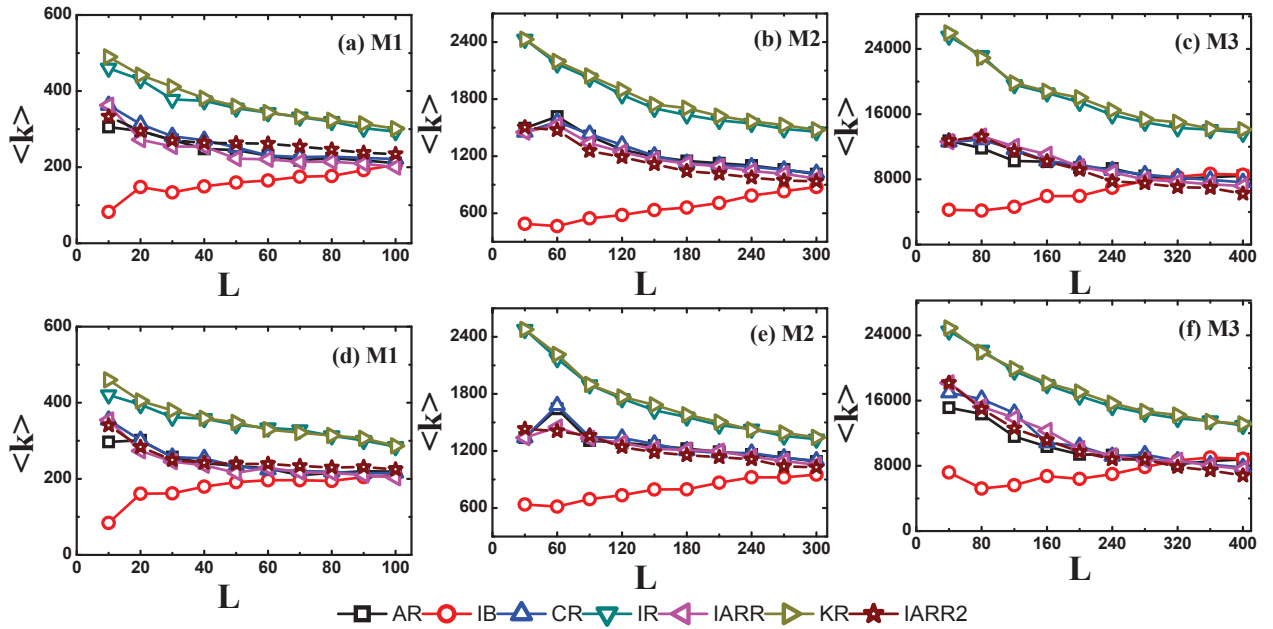}}}
\caption{\textbf{Algorithmic novelty of five methods on three MovieLens datasets.} Due to the different sizes of the three datasets, we set the maximum $L$ equal to $100$, $300$ and $400$ for M1, M2 and M3, respectively. (a)-(c) are results found by evaluation with the Academy Awards, while (d)-(f) are results by evaluation with the Golden Globe Awards.}\label{Fig5}
\end{figure}

Each method will generate a ranked list where the top-ranked movies are predicted as the nominated movies. After comparing the nominated movies that predicted right by different methods, we find that our IB method is good at finding niches (i.e., unpopular yet good movies). This ability to find novel movies is important, since finding popular movies is much easier than digging niches. Usually the niches constitute the so-called ``long tail" market which is considered to be promising and profitable. For instance, Netflix finds that in aggregate, ``unpopular" movies are rented more than popular movies, and provides a large number of niches movies on their Web site. The novelty of a movie can be measured by its degree, namely how many users have rated it.  An algorithm's novelty is defined as the average degree of the nominated movies in its ranking list, the lower the better. We compare the novelty scores of five methods. The results are shown in Fig.~\ref{Fig5}. We can see that in all presented cases, IB method always yields the lowest novelty score, indicating that IB method have higher ability to find ``dark horses" (i.e., niches, not so popular yet good movies).

Table~\ref{niches} shows the movies nominated for an Academy Award as identified by our IB method in the top-100 places, but not in the lists of other six methods in M2 dataset. The average number of ratings of the 13 movies is 455, much lower than the average number of ratings of all nominated movies in M2 dataset (i.e., 717, see table~\ref{nominate}). Besides, among the 13 movies, only three movies have been rated more than 717 times. We have also checked that the results of the other four methods highly overlapped while our IB method yields results which are considerably different from the rest. The results of other datasets are similar, so we will not present the detailed information. In the M1 dataset, there are also 27 nominated movies that are predicted right by IB method, but cannot be identified by the other four methods. The average number of ratings is 132, which is smaller than the average value of all nominated movies in the M1 dataset (i.e., 175, see table~\ref{nominate}). In the M3 dataset, there are 23 nominated movies that cannot be identified by other four methods. The average number of ratings is 3245, which is smaller than the average value of all of the nominated movies in the M3 dataset (i.e., 3942, see table~\ref{nominate}).
\begin{table}[!ht]
\centering
\caption{\textbf{The nominated movies for an Academy Award identified by IB method in the top-100 places, but not in the lists of the other six methods in the M2 dataset.} The nominated year of each movie is presented in brackets. $B$ means the movie won an Academy Award. $N$ means the movie was only nominated. Movies are ranked according to their quality scores given by the IB method.}
\begin{tabular}{lcc}
\hline
\bf{Film name} & \bf{Number of ratings}    & \bf{$N$ or $B$}     \\\hline
\textit{Apollo 13 (1995)}	&1251	&$N$ \\ 
\textit{The Apartment(1960)}	&417	&$B$ \\ 
\textit{Top Hat (1935)}	&251   &$N$ \\ 
\textit{Bonnie and Clyde (1967)}	&686	&$N$ \\ 
\textit{The Right Stuff(1983)}	&750	&$N$ \\ 
\textit{You Can't Take It With You (1938)}	&77	&$B$ \\
\textit{A Man for All Seasons(1966)}   	&219	&$B$ \\
\textit{In the Heat of the Night (1967)}	&348	&$B$ \\ 
\textit{The French Connection(1971)}	&861	&$B$ \\
\textit{Mildred Pierce (1945)}	&136&	$N$ \\ 
\textit{Mister Roberts (1955)}	&421	&$N$ \\ 
\textit{Midnight Express (1978)}	&295	&$N$ \\ 
\textit{Anatomy of a Murder (1959)} &199	&$N$ \\ \hline
\end{tabular}\label{niches}
\end{table}

\subsection{Robustness}
Besides accuracy, robustness is another important aspect to consider when selecting algorithms. Robustness usually refers to an algorithm's ability to counteract malicious activities. Here we consider the algorithm's robustness against different datasets. The intrinsic quality of an object will not change in different sampled datasets.  If an algorithm says object $A$ is better than $B$ based on sampled dataset 1, while says object $B$ is better than $A$ based on sampled dataset 2, then this algorithm is not robust because it generates inconsistent results on different sampled datasets. Therefore, instead of adding artificial ratings to investigate the algorithm's robustness, we apply MSE and the Kendall's tau~($\tau$) coefficient to measure the consistency of the results on different sampled datasets. M1, M2 and M2 are ready-made sampled datasets for experiment. Firstly, we calculate the object quality scores $Q_\alpha$ by the AR, IB, CR, IR and IARR methods in the three datasets respectively. For the three datasets, there are three pairs for comparison, namely M1 vs. M2, M2 vs. M3, and M1 vs. M3. We consider the same objects of the two datasets in each pair, and then calculate the difference between the two quality scores. $Q_\alpha^i$ and $Q_\alpha^j$ denote the quality scores of object $\alpha$ in the two datasets $i$ and $j$ ($i \neq j$), respectively, the $MSE=\frac{\sum_{i\neq j}(Q^i_\alpha-Q^j_\alpha)^2}{N_s}$, where $N_s$ is the number of same objects between datasets $i$ and $j$. The results are shown in table~\ref{RobustMSE}. In all three cases, the IB method has the lowest MSE value. Moreover, we use Kendall's tau ($\tau$) coefficient to analyze the correlation between the two ranked lists of common objects in two datasets in each pair. Table ~\ref{RobustTau} shows that the Kendall's tau ($\tau$) of the IB method is the highest among all five methods. In other words, the two ranked lists of the same objects given by the IB method in different datasets are more consistent than those given by the other four methods, indicating that IB is more robust.
\begin{table}[!ht]
\centering
\caption{\textbf{The MSE value of two ranked lists of common objects in two datasets.} For each pair of comparison, the lowest value is emphasized in bold.}
\begin{tabular}{llllll}
   \hline
   \bf{MSE}    &\bf{AR}        &\bf{IB}                &\bf{CR}           &\bf{IARR}        &\bf{IR} \\ \hline
   $M1-M2$    &0.051     &\bf{0.047}        &0.059        &0.072       &\bf{0.047}  \\ 
   $M1-M3$    &0.054     &\bf{0.052}        &0.053        &\bf{0.052}  &0.070   \\ 
   $M2-M3$    &0.022     &\bf{0.021}        &0.022         &0.024      &0.023  \\ \hline
\end{tabular}\label{RobustMSE}
\end{table}

\begin{table}[!ht]
\centering
\caption{\textbf{The kendall's tau $\tau$ correlation coefficient of two ranked lists of common objects in two datasets.} For each pair of comparison, the highest value is emphasized in bold.}
\begin{tabular}{ccccccc}
\hline
   \bf{$\tau$}      &\bf{AR}        &\bf{IB}    &\bf{CR}   &\bf{IARR}    \bf{IARR2}    &\bf{IR} \\ \hline
   $M1-M2 $   &0.753     &\bf{0.784}        &0.763        &0.781  &0.4913    &0.706  \\ 
   $M1-M3 $   &0.754     &\bf{0.791}        &0.765        &0.787  &0.5195   &0.738   \\ 
   $M2-M3 $   &0.766     &\bf{0.862}        &0.830        &0.854  &0.6475    &0.777 \\ \hline
\end{tabular}\label{RobustTau}
\end{table}

\section{Conclusions}
Building online reputation systems is important to companies who provide services or products online (i.e., Taobao e-business platform for goods~\cite{taobao}, Netflix for movies, Amazon for books/other products, Pandora for music~\cite{padora}). Since the reputation scores generated by the system's algorithm are usually used to assist users who want to buy or select something that they have no prior experience using, finding a good ranking method is important. A good method should be both effective (i.e., reflect the intrinsic values) and efficient (i.e., simple to calculate). Additionally, it must be robust against tampering. Users' rating bias greatly ruins the algorithm's performance in terms of the above three criterions. Motivated to eliminate user bias for better evaluation, we proposed an iterative balance (IB) method to identify each user's reputation and each object's quality in online rating systems. Firstly, we empirically studied the standard deviation and the skewness of users' rating scores and found that each user has a certain magnitude of rating error. Then, we introduced an equation to correct this magnitude of rating error during the iterative process. We applied mean-squared error (MSE) to measure the algorithm's accuracy on quantifying each user's reputation, and the precision to evaluate the algorithm's accuracy on identifying good objects. The algorithm's robustness is measured using both MSE and Kendall's tau coefficient. Experiments on four benchmark datasets show that the IB method is a highly self-consistent and robust algorithm. Compared with other state-of-the-art methods, the IB method has a higher ability to identify niche items (i.e., unpopular yet good objects). For example, results using the MovieLens dataset show that the IB method is good at finding the ``dark horses" for the Academy Awards. We believe our studies may find wider practical applications, such as helping online e-business platform to identify tampering, integrating the object's quality score into the recommender systems to improve the accuracy of recommendations and generally improving user experiences. Furthermore, this may also generate higher quality evaluation reports for seller reference.

\section*{Acknowledgments}

This work is supported by the National Natural Science Foundation of China (Grant No. 11622538,61673150) and the Zhejiang Provincial Natural Science Foundation of China (Grant No. LR16A050001). Z. Ren thanks the NSFC-Zhejiang Joint Fund under Grant No.U1509220.


\begin{thebibliography}{10}

\bibitem{watts2007twenty}
Watts DJ. A twenty-first century science. Nature. 2007;445(7127):489-489.

\bibitem{vespignani2009predicting}
Vespignani A. Predicting the behavior of techno-social systems. Science. 2009;325(5939):425-8.

\bibitem{RuiMao2015}
Mao R, Xu H, Wu W, Li J, Li Y, Lu M. Overcoming the challenge of variety: big data abstraction, the next evolution of data management for AAL communication systems. Communications Magazine, IEEE. 2015;53(1):42-7.

\bibitem{Alter2001}
Alter S. Information systems: foundation of e-business. 4th ed. Upper Saddle River, NJ, USA: Prentice Hall PTR; 2001.

\bibitem{lu2012recommender}
%LL, Medo M, Yeung C~H, Zhang Y~C, Zhang Z~K and Zhou T 2012 {\em Phys. Rep.\/} {\bf 519} 1.
L{\"u} L, Medo M, Yeung CH, Zhang Y-C, Zhang Z-K, Zhou T. Recommender systems. Physics Reports. 2012;519(1):1-49.

\bibitem{ren2014avoiding}
Ren X-L, L{\"u} L, Liu R, Zhang J. Avoiding congestion in recommender systems. New Journal of Physics. 2014;16(6):063057.

\bibitem{liaohao2015}
Liao H, Zeng A. Reconstructing propagation networks with temporal similarity. Scientific Reports. 2015;5:11404.

\bibitem{Resnick:2000:RS:355112.355122}
Resnick P, Kuwabara K, Zeckhauser R, Friedman E. Reputation systems. Communications of the ACM. 2000;43(12):45-8.

\bibitem{liny2016PR}
L{\"u} L, Chen D, Ren X-L, Zhang Q-M, Zhang Y-C, Zhou T. Vital nodes identification in complex networks. Physics Reports. 2016;650(1):1-63.

\bibitem{sun2012security}
Sun YL, Liu Y. Security of online reputation systems: the evolution of attacks and defenses. IEEE Signal Process Magazine. 2012;29(2):87-97.

\bibitem{liao2014Network}
Liao H, Xiao R, Cimini G, Medo M. Network-driven reputation in online scientific communities. PloS One. 2014;9(12):e112022.

\bibitem{shang2010empirical}
Shang M-S, L{\"u} L, Zhang Y-C, Zhou T. Empirical analysis of web-based user-object bipartite networks. Europhysics Letters. 2010;90(4):48006.

\bibitem{ni2014ceiling}
Ni J, Zhang Y-L, Hu Z-L, Song W-J, Hou L, Guo Q, et al. Ceiling effect of online user interests for the movies. Physica A: Statistical Mechanics and its Applications. 2014;402:134-40.

\bibitem{liu2013empirical}
Liu J, Hou L, Zhang Y-L, Song W-J, Pan X. Empirical analysis of the clustering coefficient in the user-object bipartite networks. International Journal of Modern Physics C. 2013;24(08):1350055.

\bibitem{laureti2006information}
Laureti P, Moret L, Zhang Y-C, Yu Y-K. Information filtering via iterative refinement. Europhysics Letters. 2006;75(6):1006.

\bibitem{zhou2011robust}
Zhou Y-B, Lei T, Zhou T. A robust ranking algorithm to spamming. Europhysics Letters. 2011;94(4):48002.

\bibitem{liao2014ranking}
Liao H, Zeng A, Xiao R, Ren Z-M, Chen D-B, Zhang Y-C. Ranking reputation and quality in online rating systems. PloS one. 2014;9(5):e97146.

\bibitem{whitby2004filtering}
Whitby A, J\o sang A, Indulska J. Filtering out unfair ratings in bayesian reputation systems. Proc 7th Int Workshop on Trust in Agent Societies; 2004.

\bibitem{Yu:2002:EMD:544741.544809}
Yu B, Singh MP. An evidential model of distributed reputation management. Proceedings of the First International Joint Conference on Autonomous Agents and Multiagent Systems: Part 1; Bologna, Italy. 544809: ACM; 2002;294-301.

\bibitem{brin1998anatomy}
Brin S, Page L. The anatomy of a large-scale hypertextual Web search engine. Computer Networks and ISDN Systems. 1998;30(1):107-17.

\bibitem{bharadwaj2009fuzzy}
Bharadwaj KK, Al-Shamri MYH. Fuzzy computational models for trust and reputation systems. Electronic Commerce Research and Applications. 2009;8(1):37-47.

\bibitem{Netflix}
Bennett J, Lanning S. The netflix prize. Proceedings of KDD Cup and Workshop; 2007;35.

\bibitem{MovieLens}
\url{http://www.grouplens.org}.

\bibitem{ochi2012rating}
Ochi M, Matsuo Y, Okabe M, Onai R. Rating prediction by correcting user rating bias. Web Intelligence and Intelligent Agent Technology (WI-IAT), 2012 IEEE/WIC/ACM International Conferences. 2012;1:452-6.

\bibitem{pearson1895note}
Pearson K. Note on regression and inheritance in the case of two parents. Proceedings of the Royal Society of London. 1895;58(347-352):240-2.

\bibitem{nelson2001kendall}
Nelson R. Kendall tau metric. Encyclopaedia of Mathematics. 2001;3:226-7.

\bibitem{Oscar}
The Academy Awards is commonly known as simply Oscars, is an annual American awards ceremony honoring cinematic achievements in the film industry. The awards are overseen by the Academy of Motion Picture Arts and Sciences which is a professional honorary organization. \url{http://en.wikipedia.org/wiki/Academy\_Awards}.

\bibitem{Golden}
The Golden Globe Award is an American accolade bestowed by the 93 members of the Hollywood Foreign Press Association recognizing excellence in film and television, both domestic and foreign. \url{http://en.wikipedia.org/wiki/Golden\_Globe\_Award}.

\bibitem{taobao}
Taobao is the largest e-business platform in China, which was founded by Alibaba Group in 2003. \url{http://www.taobao.com}.

\bibitem{padora}
Pandora is a website that recommend musics to users. \url{http://www.pandora.com}.

\end{thebibliography}
\end{document}